
%
%
\input uiucmac.tex

\PHYSREV
\tolerance 2000
\nopubblock
\titlepage
\title{\bf Structural Stability and Renormalization Group
for Propagating Fronts}
\author{G.C. Paquette,$^*$
Lin-Yuan Chen, Nigel Goldenfeld and Y. Oono}
\address{Department of Physics,
       Materials Research Laboratory
       and
       Beckman Institute,
       University of Illinois at Urbana-Champaign,
       1110 W.\ Green Street,
       Urbana, Il. 61801-3080}
\smallskip
\abstract
\noindent
A solution to a given equation is structurally stable if it suffers only
an infinitesimal change when the equation (not the solution) is
perturbed infinitesimally.  We have found that structural stability can
be used as a velocity selection principle for propagating fronts.
We give examples, using numerical and renormalization group methods.

\smallskip\smallskip\smallskip\noindent
$^*$ Present Address: Department of Physics, Kyoto University, Kyoto
606, Japan.

\bigskip
\rightline{P-93-08-072}

\endpage

\noindent
The steady state equation for a travelling wave propagating into an
unstable state does not always uniquely determine the wave speed.
Instead there may be multiple stable steady state travelling wave
solutions, even though the physical system described by the equation
exhibits reproducibly observable behavior corresponding to only one of
these solutions\rlap.\REFS\fisher{R.\ A.\ Fisher,  \journal Ann.
Eugenics &7&355(1937).}\REFSCON\kpp{ A.\ N.\ Kolmogorov,
I.\ G.\ Petrovskii, and N.\ S.\ Piskunov, \journal Bull.  Univ. Moskou.
Ser. Internat. Sec. A &1&1(37).}\REFSCON\Fitz{ R.\ FitzHugh, \journal J.
Biophys.  &1&445(1961); J.\ Nagumo, S.\ Arimoto, and S.\ Yoshizawa,
\journal Proc.  IRE &50&2061(1962).}\REFSCON\meinhardt{ H.\ Meinhardt,
{\sl Models for Biological Pattern Formation} (Academic Press, London,
1982).}\REFSCON\ce{P.\ Collet and J.-P.\ Eckmann, {\sl Instabilities and
Fronts in Extended Systems} (Princeton University Press, Princeton, New
Jersey, 1990).}\REFSCON\dendrite{For an early review of the dendritic
growth problem see J.S. Langer, \journal Rev. Mod. Phys. &52&1(80);
recent work is reviewed by J.S. Langer, in {\sl Chance and Matter}, J.
Souletie, J. Vannimenus, R. Stora (eds.)  (North-Holland, Amsterdam,
1987).}\refsend\  In such a situation, it is desirable to formulate a
so-called {\it selection principle}, which would allow one {\it a
priori\/} to distinguish observable from unobservable steady state front
solutions {\it without\/} having to solve directly the equation of
motion starting from the initial conditions.

For a certain class of equations, rigorous analysis shows how a wide
range of physically realizable initial conditions evolve into the
selected front, which turns out to be the slowest stable solution
allowed by the steady state equation\rlap.\Ref\AronWein{D.\ G.\ Aronson
and H.\ F.\ Weinberger, in {\it Partial Differential Equations and
Related Topics}, edited by J.\ A.\ Goldstein (Springer, Heidelberg,
1975); H.\ F.\ Weinberger, \journal SIAM J. Math. Anal. &13&3(1982); a
pedagogical introduction to this literature is given by P. Fife, {\sl
Mathematical Aspects of Reacting and Diffusing Systems}, Vol. 28 of {\sl
Lecture Notes in Biomathematics}, edited by S. Levin (Springer, New
York, 1979).}\  A physical, heuristic interpretation of this result,
known as the {\it linear marginal stability hypothesis}, has been
proposed and is believed to be applicable in the so-called {\it pulled
case}, for which the selected speed may be determined by the linear
order terms alone\rlap.\REFS\landd{J.\ S.\ Langer and
H.\ M\"{u}ller-Krumbhaar, Phys. Rev. A \journal &27&499(1983);
E.\ Ben-Jacob, H.\ R.\ Brand, G.\ Dee, L.\ Kramer, and J.\ S.\ Langer,
\journal Physica D &14&348 (1985); W.\ van Saarloos, \journal Phys.
Rev.  Lett. &58&24 (1987); W.\ van Saarloos, \journal Phys. Rev. A
&37&1(1988).}\REFSCON\dee{G.\ Dee and J.\ S.\ Langer, \journal Phys.
Rev. Lett. &50&6(1983); G.\ Dee, \journal J. Stat.  Phys.  &39&705(85);
\journal Physica D &15&295(85); G.\ Dee and W.\ van Saarloos, \journal
Phys. Rev. Lett.  &60&2641(88).}\refsend However, it is well-known that
there is another case, the so-called {\it pushed case}, where analysis
of the linear order terms alone is not sufficient to determine the
speed, and the linear marginal stability hypothesis
fails\rlap.\refmark{\landd,\dee}

The purpose of this Letter is three-fold.  First, we recall the notion
of {\it structural stability\/} --- the stability of a front with
respect to a perturbation of the governing equation --- and argue that
only structurally stable fronts are observable.  We  next show that
for structurally stable fronts, a renormalization group (RG) method can
be used to compute the change in the front speed when the governing
equation is perturbed by a marginal operator.  Finally, by combining
the structural stability principle with RG, we are able to predict the
selected front itself.  Our results apply to both the pulled and
pushed cases.  Roughly speaking, structural stability is an
insensitivity  to model modifications, whereas the RG may be interpreted
as a method to extract the structurally stable behavior of a
model\rlap.\REFS\yorg{Y.\ Oono, \journal Adv. Chem. Phys. &61&301(85);
\journal Kobunshi &28&781(79).}\REFSCON\ndgtext{N.\ D.\ Goldenfeld, {\sl
Lectures on Phase Transitions and the Renormalization Group}
(Addison-Wesley, Reading, Mass., 1992), Chapter 10.}\refsend Structural
modifications of travelling wave equations have been studied previously
(\eg, Zel'dovich's work on flame propagation\Ref\zelbar{Ya. B.
Zel'dovich, \journal Zhurn. Fiz. Khimii. &22&27(48); see also G.I.
Barenblatt, {\sl Similarity, Self-similarity, and Intermediate
Asymptotics} (Consultants Bureau, New York, 1979), p. 111.}) but to our
knowledge, structural stability has not previously been proposed as a
selection mechanism.  RG methods have previously been used to study the
asymptotics of partial differential equations
(PDEs)\refmark{\ndgtext,}\Ref\rgpde{N.\ Goldenfeld, O.\ Martin and
Y.\ Oono, \journal J. Sci.  Comp.&4&355(89); N.\ Goldenfeld, O.\ Martin,
Y.\ Oono and F.\ Liu, \journal Phys. Rev.  Lett.  &64&1361(90);
N.\ Goldenfeld and Y.\ Oono, \journal Physica A &177&213(91);
N.\ Goldenfeld, O.\ Martin and Y.\  Oono, {\sl Proceedings of the NATO
Advanced Research Workshop on Asymptotics Beyond All Orders},
S.\ Tanveer (ed.) (Plenum Press, 1992); L.\ Y.\ Chen,
N.\ D.\ Goldenfeld, and Y.\ Oono, \journal Phys. Rev. A &44&6544(91);
I.\  S.\ Ginzburg, V.\ M.\  Entov and E.\ V.\ Theodorovich, \journal J.
Appl. Maths. Mechs. &56&59(92); L.\ Y.\ Chen and N.\ D.\ Goldenfeld,
\journal Phys. Rev. A &45&5572(92); J. Bricmont, A.  Kupiainen and G.
Lin, {\sl Commun. Pure Appl. Math.} (in press).}  and propagating fronts
in the Ginzburg-Landau equation\rlap.\Ref\bricmont{J.  Bricmont and A.
Kupiainen, \journal Commun. Math. Phys. &150&193(92).}

A good model of reproducibly observable physical phenomena must give
{\em structurally stable} predictions.  That is, the observable
predictions provided by the model must be stable against \lq\lq
physically small" modifications of the system being modelled.  We will
quantify below the meaning of the term \lq\lq physically small" for a
certain class of reaction-diffusion systems.  The idea of structural
stability used here is close to that proposed by Andronov and
Pontrjagin\Ref\ap{A.\ Andronov and L.\ Pontrjagin, \journal Dokl.
Akad.  Nauk. SSSR &14&247(1937).} for dynamical systems.  In the
modeling of natural phenomena, we need not require, as did Andronov and
Pontrjagin, the structural stability of the entire model, but need only
to require it of the {\it solutions\/} corresponding to reproducibly
observable phenomena.  We call these {\it structurally stable
solutions}.  Our {\it structural stability hypothesis} states that only
structurally stable solutions of a model represent reproducibly
observable phenomena of the system being modeled.  This hypothesis is
implicit in most mathematical modeling, and indeed often redundant, yet
we will demonstrate that for reaction-diffusion equations, this
hypothesis correctly singles out observable propagating fronts.  The
basic reason for its efficacy in the situations studied here is that the
formulation of reaction-diffusion models sometimes inadvertently includes
an unphysical feature, although the model is in some sense close to a
class of physically correct models.

Consider Fisher's equation\refmark{\fisher} on the
interval $-\infty < x < \infty$:
$$
\frac{\partial \psi}{\partial t} = \frac{\partial^2\psi}{\partial x^2}
+ F ( \psi ),    \eqn\fisher
$$
where $F$ is a continuous function with $F(0) = F(1) = 0$.  We will
usually be interested in boundary conditions where $\psi$ is zero at one
boundary and unity at the other.  If $F$ satisfies the condition: $
F(\psi) > 0$ for all $\psi \in (0,1)$, then there exists a stable
travelling wave solution interpolating between $\psi = 1$ and $\psi = 0$
with propagation speed $c$ for each value of $c$ greater than or equal
to some minimum value $c^*$.  The positivity condition on $F$ stated
above together with differentiability of $F$ at the origin will
henceforth be called the AW-condition; when it is satisfied, $c^*  \ge
\hat c \equiv 2 \sqrt{F'(0)}$ .  Aronson and
Weinberger\refmark{\AronWein} proved that for \fisher\ with the
AW-condition satisfied, the selected solution is that with speed $c^*$.
In most systems studied by physicists, the minimum wave speed satisfies
$c^*=\hat c$, which corresponds to the pulled case.  Often, the initial
conditions decay sufficiently fast (faster than some exponential
function) to $\psi=0$ that the selected wave speed is in fact $c^*$.
The pushed case is equivalent to the statement $c^* > \hat c$.  In this
paper, we are concerned not only with Fisher's equation subject to the
AW condition, but with other equations or systems of equations not
satisfying the conditions required for Aronson and Weinberger's rigorous
proof, but which still exhibit the selection problem.

It is straightforward to show that all propagating solutions of
\fisher\ are structurally stable against $C^{1}$-small perturbations
$\delta F$ of $F$.  Unfortunately, reaction-diffusion equations are not
in general structurally stable with respect to $C^{0}$-small
perturbations.  Consider \fisher\ as describing the propagation of fire
along a fuse.  $F$ represents the net rate of heat production as a
function of temperature $\psi$. The value $\psi = 0$ corresponds to the
flash point, and $\psi = 1$ corresponds to the steady burning
temperature.  It is reasonable that the observable properties of such a
front would be insensitive to most small changes to $F$. However,
by altering $F$ very near $\psi = 0$ with a $C^0$-small perturbation,
$dF/d\psi$ in the neighborhood of $\psi = 0$ can be made arbitrarily
large.  That is, the rate at which heat production increases as a
function of temperature at or near the flash point can be made very
large, and this explosive low temperature behavior will travel very
rapidly along the fuse.

It is clear then that certain $C^{0}$-small perturbations are not
physically small.  This is the case, however, only for perturbations
which increase $\sup_{\psi \in \left(0,\eta \right]}(F(\psi)/\psi)$
appreciably for some $\eta >0$.  We will call a $C^{0}$-small
perturbation for which $\sup_{\psi>0} (\delta F(\psi)/\psi)$ is less
than some small positive number (which goes to zero continuously as the
$C^0$-norm of $\delta F$ vanishes) a {\em p-small}
perturbation\rlap.\Ref\footb{This condition is sufficient to insure that
the quantity
 $$\sup_{\eta >0}\left\{\sup_{\psi \in
 \left(0,\eta\right]}\left[\frac{F(\psi) + \delta F
 (\psi)}{\psi}\right] - \sup_{\psi \in
 \left(0,\eta\right]}\left[\frac{F(\psi)}{\psi}\right]\right\}$$ is
less than the same small number.}
The precise form of our structural stability hypothesis
is: physically realizable solutions of \fisher\ are those which are
stable with respect to $p$-small structural perturbations.

The ordinary differential equation (ODE) governing the travelling wave
front shape $\psi(\xi) = \psi(x,t)$ can be transformed into the equation
$$
{\dot{p}} = -c p - \frac{dU}{dq},        \eqn\party
$$
with the identifications $\xi \equiv x-ct \rightarrow t$, $\psi
\rightarrow q$, ${d \psi}/{d \xi} \rightarrow \dot{q} \equiv p$, and $F
\equiv dU/dq $.  This ODE describes the position $q$ of a unit mass
particle subject to a potential $U(q)$ and friction. The coefficient of
friction is $c$, the speed of the travelling wave.  Traveling-wave
solutions of \fisher\ interpolating between the fixed points $\psi = 0$
and $\psi = 1$ correspond in this particle analogy to trajectories which
begin at the maximum of $U$ located at $q = 1$, with zero kinetic
energy, and which terminate at the origin.  Those trajectories which
correspond to stable solutions of the original PDE are those for which
$q$ never changes sign.

If the origin is not an isolated local minimum of $U$, there is only one
value of the coefficient of friction which allows the particle to stop
here without overshooting.  If the origin is an isolated local minimum
of the potential, there is a critical value $c^*$ of the frictional
coefficient.  For all smaller values, the particle overshoots the origin
at least once, while for a continuous set of larger values, it converges
to the origin as $t \rightarrow \infty$ without overshooting.  For both
types of systems, we define $c^*$ to be the smallest value (unique value
in the former case) for which the particle approaches the origin in the
$t \rightarrow \infty$ limit without overshooting.

We have proven that $c^*$ is continuously dependent on the continuous
function $F$\rlap.\REFS\po{G.\ C.\ Paquette and Y.\ Oono,
unpublished.}\REFSCON\p{G.\ C.\ Paquette, Thesis, University of Illinois
at Urbana-Champaign, Department of Physics (1992).}\REFSCON\explain{More
precisely, for any $\epsilon >0$ there is $\delta >0$ such that for any
$p$-small $C^{0}$-perturbation, $||\delta {F}|| < \delta$ implies $
|c^{*}({F} + \delta{F}) - c^{*}({F})| < \epsilon$.  Here
$|| \;\; ||$ is the standard sup-norm, and $c^{*}({F})$ stands for
the critical frictional coefficient for the force ${F}$.}\refsend\
Thus because we can make the origin a local maximum of $U$ with
arbitrarily small $C^{0}$-perturbations (in fact, $p$-small
perturbations), and because $c^*$ is the only value of $c$ for which the
particle stops at the origin (without overshooting) in this case, the
continuity of $c^{*}$ implies that only this critical value is
structurally stable.  Therefore, our structural stability hypothesis
asserts that $c^*$ is the unique observable front propagation speed of
the original PDE.

For AW-type equations, it has been proven that $c^*$ is indeed the
selected speed. Our structural stability hypothesis is thus correct in
this case.  For non AW-type equations, however, there is no proof that
this minimum value is selected.  We have therefore performed numerical
studies to test our conclusions for non-AW equations as well as for
systems of coupled reaction-diffusion equations\rlap.\refmark{\po,\p}
An example is the equation studied by van Saarloos:  \Ref\saar{W. van
Saarloos, \journal Phys. Rev. A &39&6367(89).} $$ \frac{\pd \psi}{\pd
t}=\frac{\pd^2 \psi}{\pd x^2}-\gamma \frac{\pd^4 \psi} {\pd
x^4}+\frac{\psi}{b}(b+\psi)(1-\psi),\eqn\rgfifty $$ where
$\gamma<1/12$.  For $0<b<1/2$, the front is pushed, whereas for
$1/2<b<1$ the front is pulled.  We replaced the potential term by
$\theta(\psi - \Delta)(\psi - \Delta)(1-\psi)(\psi + b)/b$, where
$\theta$ is the step function, and let $\Delta \rightarrow 0^+$.  For
both pulled and pushed cases, the unique selected velocity converged to
that dynamically selected.

We now proceed to show how RG can be used to compute the change in the
front speed when an equation, whose structurally-stable exact solution
is known, is perturbed by a $p$-small operator.  Introducing new
variables $X \equiv e^{x}$ and $T \equiv e^{t}$, the propagating front
solution reads $\psi(x-ct) = \Phi(XT^{-c})$.  Thus the front speed is
interpreted as an anomalous dimension.  The renormalization group theory
applied to PDEs\refmark{\ndgtext,\rgpde,\bricmont} should therefore be
applicable here, too. In terms of RG, what we have found above is:
$p$-small $C^0$-perturbations are at worst marginal perturbations, but
generally, $C^{0}$-perturbations can be relevant (actually, in some
sense much worse, since the changes they produce are in certain cases
indefinitely large).

Let $\psi_{0}(x-c_0t + x_0)$ be a stable travelling front solution of
\fisher\ with speed $c_{0}$ and constant of integration $x_0$.  Let us
add a $p$-small structural perturbation $ \delta F$ to \fisher, where
its sup-norm $||\delta F||$ is of order $\epsilon$, a
small positive number\rlap,\Ref\delexp{The relationship between
$\epsilon$ and any small parameter in $\delta F$ may be complicated, as
in the example below equation \rgfifty; in many cases, $\delta F \propto
\epsilon$.}  and assume that in response the front solution is modified
to $\psi_{0} + \delta \psi$.  Defining $\xi_0\equiv x-c_0t +
x_0$ and linearizing \fisher\ with respect to $\epsilon$ in the moving
frame with velocity $c_{0}$, we formally obtain the following naive
perturbation result:
$$
\delta\psi(\xi_0,t)= e^{-c_{0}\xi_0/2}\int_{t_0}^{t}dt'
\int_{-\infty}^{+\infty}d\xi' G(\xi_0,t;\xi',t') e^{c_{0}\xi'/2}\delta
F(\psi_0(\xi')).    \eqn\naivep
$$
Here $t_{0}$ is a certain time before $\delta F(\psi_{0}(\xi_0))$
becomes nonzero, and $G$ is the Green's function satisfying
$$
\frac{\partial G}{\partial t}- {\cal L}G
=\delta(t-t')\delta(\xi-\xi') \eqn\greenf
$$
with $G \rightarrow 0$ in $|\xi - \xi'| \rightarrow \infty$, where
$$
{\cal L} \equiv \frac{\partial^{2}}{\partial \xi^{2}} +
F'(\psi_{0}(\xi)) - \frac{c_{0}^{2}}{4}. \eqn\linopp
$$
Formally, $G$ reads
$$
G(\xi,t;\xi',t') =  u_{0}(\xi)u_{0}^{*}(\xi') + \sum
e^{-\lambda_{n}(t-t')} u_{n}(\xi) u_{n}^{*}(\xi'), \eqn\greenexp
$$
where ${\cal L} u_{0} = 0$, and ${\cal L} u_{n} = \lambda_{n} u_{n}$.
The summation symbol, which may imply appropriate integration, is
over the spectrum other than the point spectrum $\{ 0 \}$.  Because the
system is translationally symmetric, $u_{0} \propto
e^{c_{0}\xi/2}\psi_{0}'(\xi)$.  Due to the known stability of the
propagating wave front, the operator ${\cal L}$ is dissipative, so 0 is
the least upper bound of its spectrum.  Hence, only $u_{0}$ contributes to
the secular term (the term proportional to $t- t_{0}$) in $\delta
\psi$.  Thus we can write
$$
 \delta \psi = -(t- t_{0}) \, \delta c \,
\psi_{0}'(\xi) + (\delta \psi)_{r},
\eqn\bare
$$
where $(\delta \psi)_{r}$ is the bounded piece (regular part), and
$$
\delta c = - \lim_{\ell \rightarrow \infty}\frac{
\int_{-\ell}^{+\ell}d\xi e^{c_{0}\xi }{\psi_{0}}'(\xi ) \delta
F(\psi_{0}(\xi ) ) }{ \int_{-\ell}^{+\ell} d\xi  e^{c_{0}\xi }
{\psi_{0}}'^{2}(\xi ) }. \eqn\dc
$$
This formula is easily justified if $0$ is isolated from the essential
spectrum of ${\cal L}$.  In the pulled case (and for some examples of
the pushed case),  this condition is not
satisfied\rlap,\refmark{\ce,}\Ref\standardtheorem{For equation
\fisher\ the essential spectrum ranges from $-\infty$ to $\max (F'(1),
F'(0) ) -c_0^2/4$.  Thus, in the pushed case, $0$ is an isolated point.}
but a more detailed argument shows that \dc\ remains
valid\rlap.\Ref\nextpr{L.\ Y. Chen, N.\ D.\ Goldenfeld and Y.\ Oono,
manuscript in preparation.}

One might immediately guess that $\delta c$ is the $O(\epsilon)$ change
in the front speed, but the naive perturbation theory is not controlled,
due to the secular divergence as $t_0\rightarrow -\infty$.  This
divergence can be controlled by perturbatively renormalizing the
constant of integration:  $x_0=x_0^R + Z$, where $x_0^R$ is the finite
observable counterpart of $x_0$, the renormalization constant
$Z=\sum_1^\infty a_n (t_0,\mu)\epsilon^n$ and the coefficients $a_n$
are chosen order by order in $\epsilon $ to eliminate the secular
divergence.  The quantity $\mu$ parameterizes the family of solutions to
the unperturbed equation; it corresponds to the arbitrary
renormalization point in the Gell-Mann---Low RG\rlap.\Ref\fngl{One may
regard our renormalization scheme as separating out the divergence by
splitting $t-t_0=t-\mu -(t_0-\mu)$, which corresponds, in the equivalent
similarity solution problem, to writing $T/e^{t_0}=(T/L)(L/e^{t_0})$.
The divergence of $\mu-t_0$ is then absorbed into $\psi$.}

To order $\epsilon$ the solution $\psi$ is given by
$$
\eqalignno{
\psi(x,t) & = \psi_0(\xi_0) -  \delta c (t-t_0) {\psi_0}' (\xi_0)
+ (\delta\psi)_r + O(\epsilon^2) & \eq \cr
& = \psi_0(\xi) + \epsilon\, a_1 {\psi_0}' (\xi) -
 \delta c (t-t_0) {\psi_0}' (\xi)
+ (\delta\psi)_r + O(\epsilon^2),  & \eq \cr
}
$$
where $\xi\equiv x - c_0t + x_0^R(\mu)$.  Thus we can choose $\epsilon
a_1= (\mu - t_0)\delta c $ to eliminate the divergence to
$O(\epsilon)$.  Requiring that $\psi$ be independent of $\mu$ gives the
RG equation
$$
\frac{\pd \psi}{\pd t}+  \delta c \frac{\pd\psi}{\pd \xi}=O(\epsilon^2).
\eqn\rgeq
$$
Thus the speed of the renormalized wave is indeed $c_{0} + \delta c$.
The formula \dc\ can also be obtained from the solvability condition for
the first order correction $\delta \psi$, and is an example of a very
general relation between renormalizability and
solvability\rlap.\refmark{\nextpr}\   Furthermore, \rgeq\ corresponds to
the amplitude equation describing the slow motion.  This relation is
also quite general\rlap.\refmark{\nextpr}

As an illustration of the use of the renormalized perturbation theory
consider the following examples.  The first, a pulled case example, is
equation \fisher\ with the nonlinear operator $F= \psi(1-\psi)$ and the
perturbation $\delta F = \epsilon \psi(1-\psi)$.  In this trivial case,
the exact result is, of course, $c^*= 2 \sqrt{1 + \epsilon}$, whereas
\dc\ gives $c^* \simeq 2 + \epsilon$.  A more interesting pushed case
example is provided by equation \rgfifty\ with $b \in (0,1/2)$.  When
$\gamma=0$, we have $c^{*}(0) = \sqrt{2b} + 1/\sqrt{2b}$.  For non-zero
$\gamma$, \dc\ gives $c^{*}(\gamma) = c^{*}(0)-\gamma
c^{*}(0)^{3}s^{4}(2 s^{2}+1)/10$ with $s \equiv 2c^{*}(0)/(c^{*}(0) +
\sqrt{c^{*}(0)^{2} - 4})$.  This agrees well with numerical
calculations. For example, this result gives $c^{*}(0.08) \simeq 2.696$
for $b=0.1$, while the corresponding value determined numerically
\refmark{\saar} is 2.715.

The perturbation theory result \dc\ can also be used to calculate
heuristically the selected speed of the unperturbed system, using the
structural stability idea.  Within perturbation theory, a necessary and
sufficient condition that $c^*$ be the selected speed is that $\delta
c(c^*)$ be bounded.  For example, when $F=\psi (1-\psi)$, the change in
the velocity $\delta c (c)$ is zero as $||\delta F||\rightarrow 0$ for
all perturbations $ \delta F$, which are both $p$-small and
differentiable at the origin, only if $c=c^*=2$; for $c>c^*$ there
exist such perturbations for which $\delta c$ does not vanish.  A simple
example of the latter is the perturbation $\delta F=\theta (u-\Delta)
(u-\Delta)(1-u) - u(1-u)$, as $\Delta\rightarrow 0^+$.

What is the physical significance of structural stability?  Returning to
the fuse analogy introduced above, we can imagine the fuse to be covered
with a very thin film of water which quickly evaporates when heated to a
temperature slightly above $\psi = 0$; nevertheless the film suppresses
tip ignition.  Thus even a small perturbation can destroy (or
drastically alter) the tip of a propagating front. For this reason, any
front whose behavior is determined by its tip can be destroyed by such a
perturbation. If and only if a front's behavior is independent of the
details of its tip, can it survive such a perturbation and be
structurally stable; hence, a front with $c > c^*$ is not structurally
stable, because no deviation is permitted from the required decay ahead
of the front.  On the other hand, the front with $c = c^*$ is
insensitive to its tip, as we can see from our explicitly dynamical RG
calculation.  There, the leading edge is determined by the initial
conditions, is not universal, and vanishes as $t\rightarrow\infty$.
Nevertheless, for sufficiently rapid leading edge decay in space, the
asymptotic speed is $c^*$: thus $c^*$ is independent of the details of
the leading edge, so that the selected front is structurally stable.

\ACK
The authors are grateful to Paul Newton for valuable discussions.  This
work is, in part, supported by the National Science Foundation Grant
NSF-DMR-90-15791.  LYC acknowledges the support of grant
NSF-DMR-89-20538, administered through the University of Illinois
Materials Research Laboratory. GCP acknowledges support from
the Japanese Society for the Promotion of Science.

\vfill\eject

\refout\end